\documentclass[reprint,aps,pra,superscriptaddress]{revtex4-2}
\usepackage{graphicx}
\usepackage{xifthen}
\usepackage{xspace}
\usepackage{amsmath}
\usepackage{amssymb}
\usepackage{comment}
\usepackage[colorlinks, allcolors=blue]{hyperref}
\usepackage[all]{hypcap}
\usepackage[mathlines]{lineno}
\usepackage{braket}
\usepackage{wrapfig}
\usepackage{lipsum}
\usepackage{ulem}
\usepackage{units}
\usepackage{xcolor}
\usepackage[utf8]{inputenc}
\begin{document}

\title{Performance analysis of different photon-mediated entanglement generation schemes under optical dephasing and spectral diffusion}% Force line breaks with \\
\author{Kinfung Ngan}
\author{Shuo Sun}
\email{shuosun@colorado.edu}
\affiliation{
JILA and Department of Physics, University of Colorado Boulder, Colorado 80309, USA
}

\date{\today}% It is always \today, today,

\begin{abstract}
Solid-state quantum emitters, such as quantum dots, color centers, rare-earth dopants, and organic molecules, offer qubit systems that integrate well with chip-scale photonic and electronic devices. To fully harness their potential for quantum applications requires the generation of entanglement between two remote qubits with high fidelity and efficiency. In this article, we compare the performance of three common photon-mediated entanglement schemes under realistic noise for solid-state quantum emitters, including optical dephasing and spectral diffusion. We identify the optimal scheme across different noise regimes and calculate the measurement parameters needed to achieve the highest entanglement fidelity at a given rate. Additionally, we explore the effects of temporal and spectral filtering in enhancing entanglement fidelity. Our findings provide practical guidelines for selecting optimal entanglement schemes and outline the measurement strategies for achieving better entanglement fidelity.

\end{abstract}

\maketitle

\section{Introduction}
\label{intro}
%\tableofcontents
Solid-state quantum emitters such as quantum dots \cite{lodahl2015interfacing}, color centers \cite{atature2018material,janitz2020cavity,bradac2019quantum}, rare-earth dopants \cite{zhong2019emerging,zhou2023photonic}, and organic molecules \cite{toninelli2021single}, are trap-free qubit systems that can be directly integrated with chip-scale photonic and electronic devices \cite{pelucchi2022potential}. To truly harness the quantum advantages of these systems requires the generation of high-fidelity entanglement between two remote qubits at a fast rate. Such remote entanglement is essential for extending the scale of spin-based quantum information processors \cite{nemoto2014photonic,simmons2024scalable}, entanglement-enhanced quantum sensing \cite{degen2017quantum,zhang2021distributed}, and the realization of quantum repeaters and quantum networks \cite{wehner2018quantum,azuma2023quantum}.

Optical photons are ideal carriers to mediate remote entanglement. They are highly versatile interconnects and can bridge quantum interactions over multiple distance scales, from microns to kilometers, to form complex distributed quantum systems. Several schemes have been proposed for photon-mediated entanglement generation \cite{beukers2024remote, wein2020analyzing}. These schemes all rely on the generation of spin-tagged photons by each emitter, followed by the erasure of the photon's which-path information via the detection of photons mixed on a beamsplitter. The major difference between these schemes is how the spin-tagged photons are generated. To date, three physical mechanisms for generating spin-tagged photons have been proposed and experimentally demonstrated on solid-state spin systems, including spontaneous emission \cite{bernien2013heralded,humphreys2018deterministic,ruskuc2024scalable}, Raman emission \cite{delteil2016generation, stockill2017phase, sipahigil2016integrated}, and resonant scattering \cite{knaut2024entanglement}. However, it is unclear which mechanism offers the best performance in entanglement generation for solid-state spins. In particular, given that many solid-state quantum emitters suffer from optical dephasing and spectral diffusion caused by their environment, it remains an open question how different entanglement generation schemes behave under different realistic noises, and which scheme is optimal.

In this article, we quantitatively compare the performance of three different photon-mediated entanglement generation schemes under optical dephasing and spectral diffusion, and identify the optimal scheme in different noise parameter regimes. For each scheme, we calculate the optimal measurement parameters needed to achieve the best entanglement fidelity at a targeted entanglement rate. In addition, we compare the performance of temporal and spectral filtering in boosting the entanglement fidelity.  Our analysis provides a clear guide on the selection of the optimal photon-mediated entanglement generation schemes for different solid-state quantum emitters, and lays out the parameter requirements for future experimental realizations of different schemes.

 This article is organized as follows. In Sec. \ref{Protocal review}, we review the three different schemes for photon-mediated entanglement generation. In Sec. \ref{method}, we define the entanglement fidelity and efficiency, which we use as figures of merit to quantify the performance of the entanglement generation process. We also present how we calculate these figures of merit for each entanglement generation scheme. Sec. \ref{filtering} studies the effectiveness of temporal and spectral filtering in boosting the entanglement fidelity. In Sec. \ref{comparison}, we compare the optimal entanglement fidelity of the three different schemes under optical dephasing and spectral diffusion. Sec. \ref{conclusion} concludes the paper.

\section{Review of photon-mediated entanglement generation schemes} 
\label{Protocal review}

In this section, we review three schemes for photon-mediated entanglement generation, where the spin-tagged photons are generated by spontaneous emission, Raman emission, and resonant scattering, respectively. For each scheme, the entanglement can be heralded by either the detection of a single photon or two photons. To ensure a fair comparison of the performance, we will focus on only the single-photon heralding version of each scheme. This ensures that the relative entanglement generation efficiency among different schemes only depends on the physical process of the entanglement generation, and is independent of the photon collection efficiency of the experimental apparatus. Our analysis holds true qualitatively if one were to compare the performance of the two-photon heralding versions of each scheme.

\begin{figure}[t]
     \centering
         \includegraphics[width=0.48\textwidth]{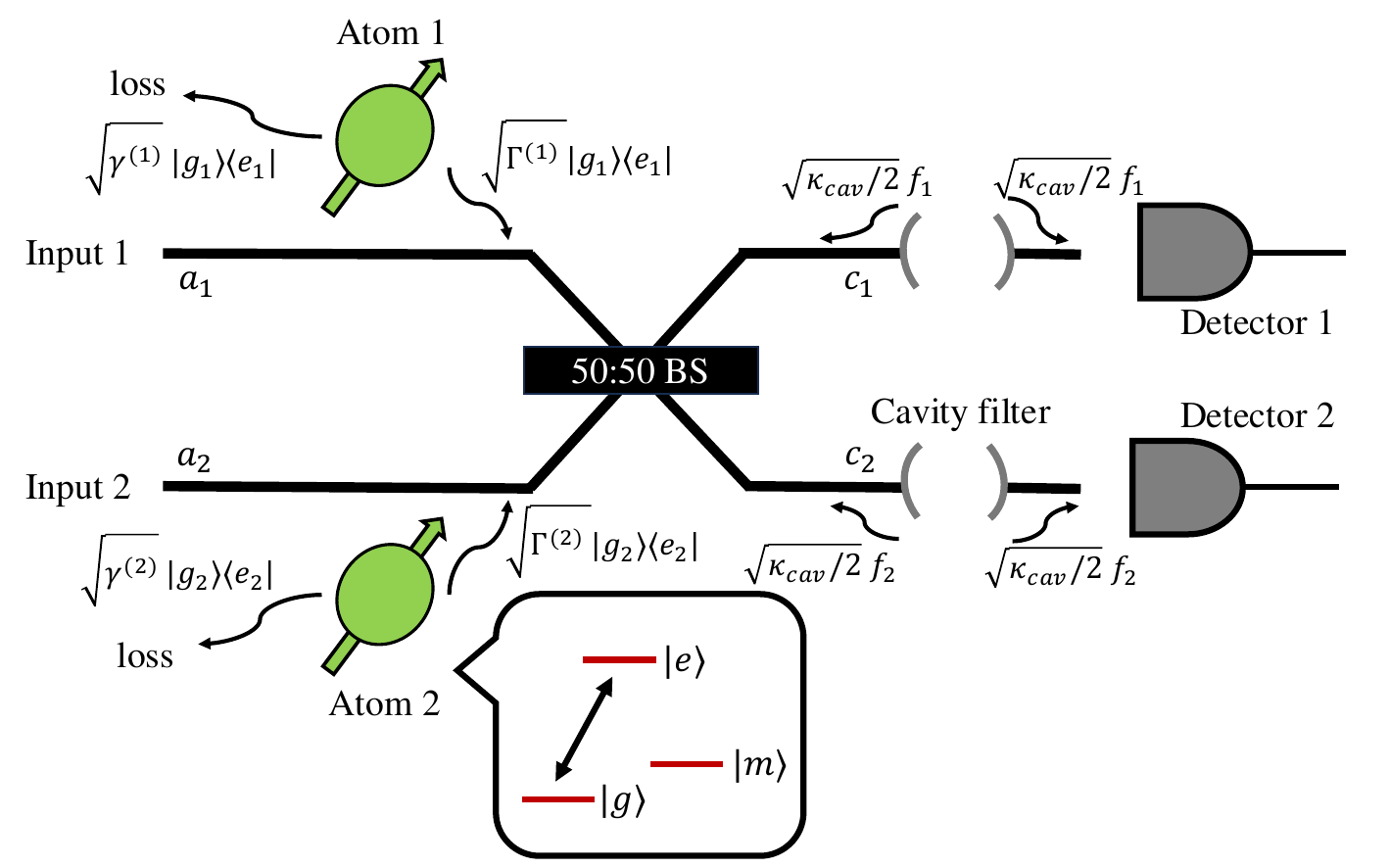}
         \caption{ Schematic setup for photon-mediated entanglement generation between two remote atoms. Spin-tagged photons from each atom are first coupled to the waveguide mode $a_1$ and $a_2$ respectively, and then mixed on a 50/50 beamsplitter. A single photon detection event on either detector 1 or 2 at the output of the beamsplitter will herald the two atoms into an entangled state. To account for spectral filtering, we have included a cavity right before each detector. The inset shows the schematic energy levels of the atom being considered in all the entanglement generation schemes.  }
         \label{SP}
\end{figure}

Figure \ref{SP} depicts a schematic setup we consider for photon-mediated entanglement generation. We assume each atom contains a $\lambda$-type energy level structure with two ground states $\ket{g_k}$ and $\ket{m_k}$, and an optically excited state $\ket{e_k}$ (where $k\in \{1,2\}$ identifies the atoms). In each atom, the two ground states form a stable spin qubit, and the excited state enables optical transitions that will be used for photon-mediated entanglement generation. We assume that the transitions between the excited state $\ket{e_k}$ and the metastable ground state $\ket{m_k}$ are significantly weaker than the transitions between $\ket{e_k}$ and $\ket{g_k}$, such that we can ignore spontaneous emission from $\ket{e_k}$ to $\ket{m_k}$. Such a spin-photon interface has been realized in many optically active solid-state quantum emitters, including quantum dots \cite{sun2016quantum, sun2018single}, color centers \cite{nguyen2019quantum}, and rare-earth ions \cite{kindem2020control, zhong2018optically, chen2020parallel}. To generate photon-mediated entanglement, we use a waveguide to collect spin-tagged photons from each atom. These spin-tagged photons may be generated via spontaneous emission, Raman emission, or resonant scattering, which we will describe separately below. We assume that the transition $\ket{g_k} \leftrightarrow \ket{e_k}$ of each atom couples to the waveguide $k$ with a spontaneous emission rate $\Gamma^{(k)}$ while decaying to all other modes with a spontaneous emission rate $\gamma^{(k)}$. We use a balanced beamsplitter to erase the which-path information of the collected photons, which implements the following unitary transformation
\begin{equation}
  \begin{pmatrix}
        c_{1} \\
        c_{2} 
\end{pmatrix}
= \frac{1}{\sqrt{2}}
\begin{pmatrix}
        1 & -1 \\
        1 & 1
\end{pmatrix}
\begin{pmatrix}
        a_{1} \\
        a_{2} 
\end{pmatrix},
\end{equation}
where $a_{k}$ is the photon annihilation operator of the input waveguide mode $k$, and $c_{k}$ is the photon annihilation operator of the output waveguide mode $k$.  We then herald the entanglement by detecting the spin-tagged photons using one of the single-photon detectors placed in the mode $c_1$ or $c_2$. To closely match experimental conditions, we assume the single-photon detectors are binary and cannot resolve photon numbers.

In our most general model, we also include temporal and spectral filtering, two techniques that can be used to boost the entanglement fidelity. To include temporal filtering, we assume that the detectors have a time resolution much shorter than the atomic spontaneous emission lifetime, such that we can herald the entanglement based on photons detected within an arbitrary time window as determined by the user. To include spectral filtering, we assume that the spin-tagged photons in the modes $c_1$ and $c_2$ are critically coupled to a cavity with a decay rate $\kappa_{cav}$ before detection. 

\textit{Spontaneous Emission Scheme}. This scheme generates spin-tagged photons via spontaneous emission. First, we prepare both atoms in an unbalanced superposition state $\alpha \ket{g_{k}}+\beta \ket{m_{k}}$, where $\alpha$ and $\beta$ are arbitrary complex numbers satisfying $|\alpha|^2+|\beta|^2 = 1$ and $|\alpha|\ll 1$. We then apply an optical $\pi$ pulse that drives the transition $\ket{g_{k}} \leftrightarrow \ket{e_{k}}$, which transforms the state of the two atoms into $\ket{\psi} = \alpha^{2}\ket{e_{1}e_{2}}+\beta^{2}\ket{m_{1}m_{2}}+\alpha\beta(\ket{e_{1}m_{2}}+\ket{m_{1}e_{2}})$. Upon spontaneous emission of each atom, the state of the joint system becomes $\ket{\psi} = \alpha^{2}\ket{g_{1}g_{2}}a^{\dagger}_{1}a^{\dagger}_{2}\ket{vac}+\beta^{2}\ket{m_{1}m_{2}}\ket{vac}+\alpha\beta(\ket{g_{1}m_{2}}a^{\dagger}_{1}\ket{vac}+\ket{m_{1}g_{2}}a^{\dagger}_{2}\ket{vac})$. Note that here we have neglected the terms containing one or more photons in the lossy modes for simplicity. These terms will be naturally eliminated in the end when we focus on the atomic state conditioned on a detector click. For the same reason, we can safely drop the second term $\beta^{2}\ket{m_{1}m_{2}}\ket{vac}$. In addition, we can neglect the first term $\alpha^{2}\ket{g_{1}g_{2}}a^{\dagger}_{1}a^{\dagger}_{2}\ket{vac}$ since the coefficient $\alpha^2$ is much smaller than the coefficients of the other three terms. Under these assumptions, we can write the unnormalized state of the joint system as $\ket{\psi} = \alpha\beta(\ket{g_{1}m_{2}}a^{\dagger}_{1}\ket{vac}+\ket{m_{1}g_{2}}a^{\dagger}_{2}\ket{vac})$. Rewriting this state in terms of the output modes of the beamsplitter, we have the unnormalized state $ \ket{\psi} = \alpha\beta( \ket{\Psi^{-}}c^{\dagger}_{1}\ket{vac}+\ket{\Psi^{+}}c^{\dagger}_{2}\ket{vac})$, where $\ket{\Psi^{\pm}} = \frac{1}{\sqrt{2}}(\ket{g_{1}m_{2}} \pm \ket{m_{1}g_{2}})$ is the ideal Bell state. Therefore, the state of the two atoms will be projected into an ideal Bell state $\ket{\Psi^{-}}$ or $\ket{\Psi^{+}}$ conditioned on a click on either the detector 1 or detector 2.

 \textit{Raman emission scheme}. In this scheme, the spin-tagged photons are generated by Raman emission. We start by initializing both atoms in state $\ket{m_{k}}$. We then apply an optical pulse that drives the transition $\ket{e_{k}} \leftrightarrow \ket{m_k}$ of each atom with a Rabi frequency $\Omega_k$ and a frequency detuning $\delta_k $. When $\Omega_k \ll \delta_k$, the optical pulse will probabilistically generate a Raman scattered photon (with a probability of $p_k$) while flipping the atomic ground state from $\ket{m_k}$ to $\ket{g_{k}}$, without populating the excited state. By adjusting the optical pulse duration, the detuning $\delta_k$, and the Rabi frequency $\Omega_k$, we can make the Raman emission probabilities for the two atoms the same (i.e. $p_1 = p_2 = p$), while satisfying $p \ll 1$. Under this condition, the joint state of the system upon the optical drive can be written as $\ket{\psi} = p\ket{g_{1}g_{2}}a^{\dagger}_{1}a^{\dagger}_{2}\ket{vac} + \sqrt{p(1-p)}\ket{g_{1}m_{2}}a^{\dagger}_{1}\ket{vac} + \sqrt{p(1-p)}\ket{m_{1}g_{2}}a^{\dagger}_{2}\ket{vac} +  (1-p)\ket{m_{1}m_{2}}\ket{vac}$, where we again ignored the term corresponding to a photon scattered into the lossy modes.  Ignoring the first term (since $p$ is small) and the last term (since it will not result in the detection of a photon), and rewriting the state in the output modes of the beamsplitter, we can again express the unnormalized joint state as $\ket{\psi} =  \sqrt{p(1-p)}(\ket{\Psi^{-}} c^{\dagger}_{1}\ket{vac}+\ket{\Psi^{+}} c^{\dagger}_{2}\ket{vac})$. Therefore, the state of the two atoms will be projected into an ideal Bell state $\ket{\Psi^{-}}$ or $\ket{\Psi^{+}}$ conditioned on a click on either the detector 1 or detector 2. 

\textit{Resonant scattering scheme}. This scheme generates spin-tagged photons via resonant scattering of photons from the atom-waveguide coupled system. Initially, we prepare both atoms into the balanced superposition state $(\ket{g_{k}} +\ket{m_{k}})/\sqrt{2}$, and prepare a single photon (typically realized by a weak coherent state) into a balanced superposition state $\frac{1}{\sqrt{2}}(a^{\dagger}_{1} +a^{\dagger}_{2})\ket{vac}$ as well. The input photon is resonant with the optical transition $\ket{g_k} \leftrightarrow \ket{e_k}$ of each atom. The quantum state of the joint system before photon scattering becomes $\ket{\psi} = \frac{1}{2}(\ket{g_{1}g_{2}} +\ket{m_{1}m_{2}} +\ket{g_{1}m_{2}} +\ket{m_{1}g_{2}})\frac{1}{\sqrt{2}}(a^{\dagger}_{1} +a^{\dagger}_{2})\ket{vac}$. At the large cooperativity regime where $\Gamma^{(k)} \gg \gamma^{(k)}$, each atom applies a $\pi$-phase shift on the photon if it is in the state $\ket{g_k}$ \cite{waks2006dipole, shen2009theory}, but no phase shift if it is in the state $\ket{m_k}$. The joint state of the system becomes $\ket{\psi} = \frac{-1}{2\sqrt{2}}(\ket{g_{1}g_{2}}(a^{\dagger}_{1} +a^{\dagger}_{2}) \ket{vac}
+ \ket{g_{1}m_{2}}(a^{\dagger}_{1}-a^{\dagger}_{2})\ket{vac}
- \ket{m_{1}g_{2}}(a^{\dagger}_{1} - a^{\dagger}_{2})\ket{vac}
- \ket{m_{1}m_{2}}(a^{\dagger}_{1}+a^{\dagger}_{2})\ket{vac}
)$. Expressing this state in terms of the output modes of the beamsplitter, the joint state of the system becomes $\ket{\psi} = \frac{-1}{\sqrt{2}}(\ket{\Psi^{-}}c^{\dagger}_{1}\ket{vac} + \ket{\Phi^{-}}c^{\dagger}_{2}\ket{vac})$, where $\ket{\Phi^{-}} = \frac{1}{\sqrt{2}}(\ket{g_{1}g_{2}} - \ket{m_{1}m_{2}})$. Conditioned on a click on either detector, the state of the two atoms will collapse to the corresponding Bell state.

\section{Theory framework} 
\label{method}
In this section, we present the theory framework for calculating entanglement fidelity and efficiency, two figures of merit that we use for characterizing the performance of the entanglement generation schemes. Without the loss of generality, we focus our analysis on the entangled state heralded by a click on detector 1, where the targeted entangled state is $\ket{\Psi^-}$. We can then define the entanglement fidelity $F$ as 
\begin{equation}\label{expression of F}
    F = \frac{\bra{\Psi^{-}}\rho_{c}(T)\ket{\Psi^{-}}}{tr[\rho_{c}(T)]}, 
\end{equation}
where $\rho_{c}(T)$ is the conditional density matrix of the two atoms at time $T$, conditioned on a click at detector 1 within a time window from $t=0$ to $t=T$. We define the entanglement efficiency as 
\begin{equation}
    \eta = tr[\rho_{c}(T)],
\end{equation}
which measures the probability of registering a click at detector 1 within the time window from $t=0$ to $t=T$ in each entanglement generation attempt. Here we implicitly assume a photon collection efficiency of 1. A non-unity collection and detection efficiency will degrade the entanglement efficiency of each scheme by a fixed factor, but it will not change the relative efficiency among the three schemes. We also assume that the temporal filtering of the spin-tagged photon always employs a time window that starts at $t = 0$. This is because this setting ensures the largest entanglement efficiency given the same entanglement fidelity.

We calculate the conditional density matrix $\rho_{c}(T)$ as $\rho_{c}(T) = \rho_{r}(T) - \rho_{null}(T)$, where $\rho_{r}(T)$ is the density matrix of the two atoms at time $T$ without conditioning on any photon detection, and $\rho_{null}(T)$ is the density matrix of the atoms at time $T$ conditioned on no click at detector 1 within the time window from $t=0$ to $t=T$ (i.e. no photon emitted into the mode $c_1$ within this time window). The unconditional density matrix $\rho_{r}(T)$ can be calculated by solving the standard master equation,
\begin{equation}\label{Lindblad equation}
\begin{aligned}
\dot{\rho_r} = \mathcal{L}\rho_r
= -i[H_{sys},\rho_r] + \sum_{k=1,2} \{ D[L_{wg}^{(k)}](\rho_r) +\\ D[L_{loss}^{(k)}](\rho_r) + D[L_{dp}^{(k)}](\rho_r)\},
\end{aligned}
\end{equation}
where $\mathcal{L}$ is the Liouvillian superoperator that governs all the coherent and incoherent dynamics of the atomic system, $H_{sys}$ is the Hamiltonian of the system, and $D[O]$ is the Lindblad superoperator that operates as $D[O](\rho) = O\rho O^{\dagger} - \frac{1}{2}(O^{\dagger}O\rho + \rho O^{\dagger}O)$, and $L_{wg}^{(k)}$, $L_{loss}^{(k)}$, and $L_{dp}^{(k)}$ describe the dissipation of the $k$-th atom into the waveguide mode $a_{k}$,  the dissipation of the $k$-th atom into the photonic leaky modes, and the optical dephasing of the $k$-th atom respectively, given by $L_{wg}^{(k)} = \sqrt{\Gamma^{(k)}} \ket{g_{k}}\bra{e_{k}}$, $L_{loss}^{(k)} = \sqrt{\gamma^{(k)}}\ket{g_{k}}\bra{e_{k}}$, and $L_{dp}^{(k)} = \sqrt{2\gamma^{(k)}_{dp}}\ket{e_{k}}\bra{e_{k}}$, where $2\gamma_{dp}^{(k)}$ is the homogeneously broadened optical linewidth of the $k$-th atom due to optical dephasing.
The density matrix $\rho_{null}(T)$ can be calculated by solving the revised master equation
\begin{equation}\label{Lindblad equation 2}
\dot{\rho}_{null} = (\mathcal{L}\rho_{null} - c_{1}\rho_{null}c_{1}^{\dagger}),
\end{equation}
where $c_{k}= \frac{1}{\sqrt{2}}(L^{(1)}_{wg} +(-1)^{k} L^{(2)}_{wg})$ is the jump operator corresponding the emission of a photon into the $k$-th output waveguide mode.
The system Hamiltonian $H_{sys}$ depends on the exact entanglement scheme we are considering. For the spontaneous emission scheme, we consider that the system of the two atoms is initially prepared in the state $\ket{\psi} = (\alpha\ket{e_{1}}+\beta\ket{m_{1}})(\alpha\ket{e_{2}}+\beta\ket{m_{2}})$ (i.e., we do not model the physical process of the optical $\pi$ pulse drive). In this case, the system Hamiltonian takes a simple form, given by 
\begin{equation}\label{spontaneous H}
    H_{sys} = \sum_{k=1,2}\delta_{k}\ket{e_{k}}\bra{e_{k}},
\end{equation}
where $\delta_{k} = \omega_k - \omega_{ref}$ is the frequency difference between the atomic transition $\ket{g_k} \leftrightarrow \ket{e_k}$ and a reference frequency, which can be arbitrary in this scheme. For the Raman emission scheme, the system Hamiltonian in the rotating frame is given by
\begin{equation}\label{Raman H}
H_{sys} = \sum_{k=1,2} \delta_{k}\ket{e_{k}}\bra{e_{k}} + \Omega_{k}(t)(\ket{e_{k}}\bra{m_{k}} + \ket{m_{k}}\bra{e_{k}}),
\end{equation}
where $\Omega_{k}(t)$ is the Rabi frequency that describes the classical drive. The frequency reference $\omega_{ref}$ is set to the frequency of the classical drive $\omega_{d}$. For the resonant scattering scheme, we consider that the incident optical field is in a weak coherent state $\ket{\beta(t)}$ in both input waveguide modes. Under this condition, the system dynamics are governed by the following Hamiltonian derived from Mollow transformation or the SLH formalism \cite{combes2017slh},
\begin{equation}\label{resonant H}
\begin{gathered}
\begin{aligned}
 H_{sys} &= \sum_{k=1,2} \{ \delta_{k}\ket{e_k}\bra{e_k} + \\
&\frac{\sqrt{\Gamma^{(k)}}}{2i}(\beta(t)\ket{e_{k}}\bra{g_{k}} - \beta^{*}(t)\ket{g_{k}}\bra{e_{k}} ) \}.
\end{aligned}
\end{gathered}
\end{equation}
Here, the reference frequency $\omega_{ref}$ is given by the frequency of the incident coherent field. Following the derivation, the jump operator $L_{wg}^{(k)}$ will be redefined as $L_{wg}^{(k)} = \sqrt{\Gamma^{(k)}}\ket{g_k}\bra{e_k} + \beta(t)$.

 To account for spectral filtering, we assume that the spin-tagged photons in the modes $c_1$ and $c_2$ are critically coupled to a cavity before detection. We model the cascaded system using the SLH formalism \cite{combes2017slh}. The density matrix $\rho_r$ will evolve by $\dot{\rho_r} = \mathcal{L}_{new}\rho_r$, where $\mathcal{L}_{new}$ is given by
\begin{equation}
\begin{gathered}
\begin{aligned}
 \mathcal{L}_{new}\rho_{r} = -i[H_{sys} + H_{cav},\rho_r]   + \sum_{k=1,2} \{ D[L_{t}^{(k)}](\rho_r) +\\ D[L_{r}^{(k)}](\rho_r)+D[L_{loss}^{(k)}](\rho_r) + D[L_{dp}^{(k)}](\rho_r)\}, 
\end{aligned}
\end{gathered}
\end{equation}
where $H_{cav}$ is the Hamiltonian describing the interaction between the waveguide mode and the filter cavity, given by
\begin{equation}
\begin{gathered}
\begin{aligned}
 H_{cav} &= \sum_{k=1,2} \frac{1}{2i}\sqrt{\frac{\kappa_{cav}}{2}}(f^{\dagger}_{k}c_{k} - f_{k}c_{k}^{\dagger}),
\end{aligned}
\end{gathered}
\end{equation}
and $L_t^{(k)}$ and $L_r^{(k)}$ describes the dissipation of the filter cavity into the transmitted mode and the reflected mode, respectively,  given by  $L_t^{(k)} = \sqrt{\frac{\kappa_{cav}}{2}}f_{k}$, and $L_r^{(k)} = c_{k} + \sqrt{\frac{\kappa_{cav}}{2}}f_{k}$, and $f_{k}$  is the cavity mode annihilation operator of the $k$-th cavity. The density matrix $\rho_{null}$ will evolve according to
\begin{equation}\label{Lindblad equation 3}
\dot{\rho}_{null} = (\mathcal{L}_{new}\rho_{null} - L_{t}^{(1)}\rho_{null}L_{t}^{(1)\dagger}).
\end{equation}

The above theoretical framework allows us to calculate the entanglement fidelity and efficiency when the atoms are stationary. Many solid-state artificial atoms suffer from spectral diffusion \cite{sun2016quantum,ngan2023quantum,ruskuc2024scalable}, where the atomic optical transition frequency slowly fluctuates due to fluctuations in the bath charges, external magnetic field, strain, etc. Different from optical dephasing, the timescale of spectral diffusion is typically much slower than each individual entanglement attempt. However, the timescale of spectral diffusion is typically much faster than the total time of an experiment where a large number of entanglement attempts have to be made in order to evaluate the statistically averaged entanglement fidelity. Here, we assume that both atoms have the same center frequencies for their atomic optical transitions, and model the spectral diffusion by computing the entanglement fidelity and efficiency based on an ensemble average of different frequency detunings, given by 
\begin{equation}\label{avg F}
    \bar{F} = \int^{\infty}_{-\infty} \int^{\infty}_{-\infty} F (\delta_1, \delta_2)f(\delta_{1},\delta_{2},\xi_{1},\xi_{2} ) d\delta_{1}d\delta_{2},
\end{equation}
\begin{equation}\label{avg eta}
    \bar{\eta} = \int^{\infty}_{-\infty} \int^{\infty}_{-\infty} \eta(\delta_1, \delta_2)f(\delta_{1},\delta_{2},\xi_{1},\xi_{2} ) d\delta_{1}d\delta_{2},
\end{equation}
where $f(\delta_{1}, \delta_{2}, \xi_{1}, \xi_{2})$ represents an uncorrelated bivariate normal distribution, expressed as 
\begin{equation}\label{binormal distribution}
f(\delta_{1}, \delta_{2}, \xi_{1}, \xi_{2}) = \frac{1}{ 2\pi \xi_{1}\xi_{2}}e^{-\frac{1}{2}\frac{\delta_{1}^{2}}{\xi_{1}^{2}}}
e^{-\frac{1}{2}\frac{\delta_{2}^{2}}{\xi_{2}^{2}}}
\end{equation}
Here, $\xi_{k}$ represents the standard deviation associated with the spectral diffusion linewidth $\gamma_{sd}^{(k)}$, given by $\xi_{k} = 2.35\gamma_{sd}^{(k)}$. 

\section{Temporal and Spectral Filtering}
\label{filtering}
\begin{figure}[t!]
     \centering
         \includegraphics[width=0.5\textwidth]{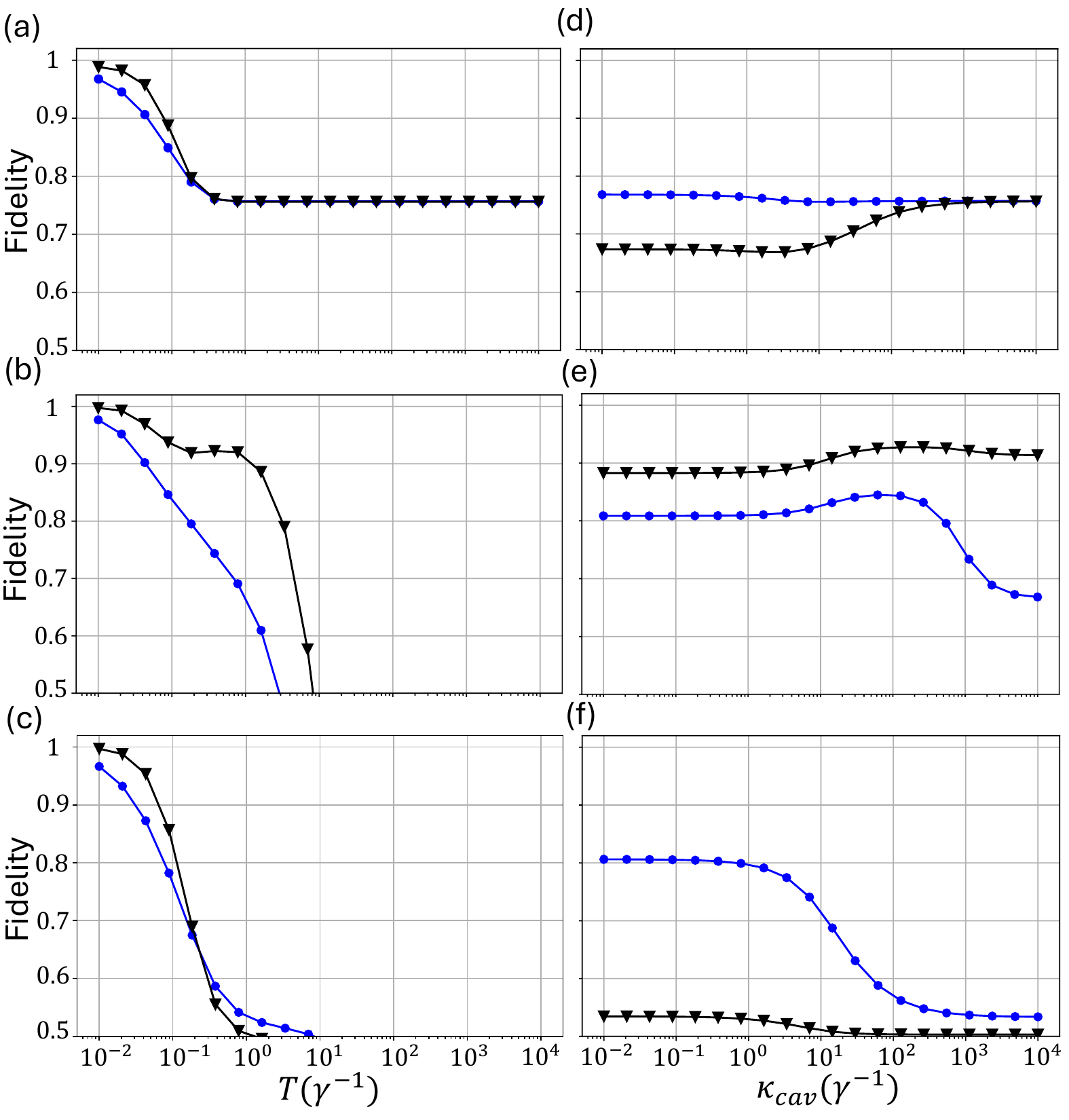}
         \caption{(a) - (c) Entanglement fidelity as a function of the integration time window for the spontaneous emission scheme (a), the Raman emission scheme (b), and the resonant scattering scheme (c). (d) - (f) Entanglement fidelity as a function of filter bandwidth for the spontaneous emission scheme (d), the Raman emission scheme (e), and the resonant scattering scheme (f). For all panels, the blue circles represent the fidelities when only optical dephasing is included, and black triangles represent the fidelities when only spectral diffusion is included.} 
         \label{filter}
\end{figure}

Both optical dephasing and spectral diffusion degrade the entanglement fidelity, as they reduce the indistinguishability of spin-tagged photons from the two emitters, regardless of the entanglement schemes we employ. One may expect that we can simply boost the entanglement fidelity by temporal or spectral filtering of the spin-tagged photons. In this section, we will discuss the effects and limitations of temporal and spectral filtering in boosting the entanglement fidelity.

We first focus on temporal filtering. Figure \ref{filter} (a) - (c) show the calculated entanglement fidelity as a function of the integration time window $T$, for the spontaneous emission scheme (a), the Raman emission scheme (b), and the resonant scattering scheme (c), respectively. The blue circles represent the entanglement fidelity when only optical dephasing is present ($\gamma^{(1)}_{dp} = \gamma^{(2)}_{dp} = 5\gamma $ and $\gamma^{(1)}_{sd} = \gamma^{(2)}_{sd} = 0$). The black triangles represent the entanglement fidelity when only the spectral diffusion is present ($\gamma^{(1)}_{sd} = \gamma^{(2)}_{sd} = 5\gamma $ and $\gamma^{(1)}_{dp} = \gamma^{(2)}_{dp} = 0$). We set the rest of the parameters to be $\gamma^{(1)} = \gamma^{(2)} = \gamma $, $\Gamma^{(1)} = \Gamma^{(2)} = 10\gamma $, $\alpha = 0.1$ (for the spontaneous emission scheme only),  $
\Omega= 60\gamma$ and $ \delta_{1} = \delta_{2} = \Delta = 600 \gamma $  (for the Raman emission scheme only), and $\beta = \sqrt{0.01\gamma}$ (for the resonant scattering scheme only).

For all three schemes, we observe that the entanglement fidelity increases to unity when a sufficiently narrow time window is applied, and temporal filtering proves effective in restoring the fidelity degraded by both optical dephasing and spectral diffusion. This conclusion can be understood intuitively. For optical dephasing, a shorter time window reduces the probability that a phase jump occurs during spontaneous emission, Raman emission, or resonant scattering, thereby boosting the entanglement fidelity. For spectral diffusion, a shorter time window effectively erases the which-path information leaked by the differences in the atomic frequencies, thereby boosting the entanglement fidelity.
 
We then move on to the effect of spectral filtering. Figure \ref{filter} (d) - (f) show the calculated entanglement fidelity as a function of the spectral filter bandwidth $\kappa_{cav}$, for the spontaneous emission scheme (d), the Raman emission scheme (e), and the resonant scattering scheme (f), respectively. The blue circles represent the entanglement fidelity when only optical dephasing is present ($\gamma^{(1)}_{dp} = \gamma^{(2)}_{dp} = 5\gamma $ and $\gamma^{(1)}_{sd} = \gamma^{(2)}_{sd} = 0$). The black triangles represent the entanglement fidelity when only the spectral diffusion is present ($\gamma^{(1)}_{sd} = \gamma^{(2)}_{sd} = 5\gamma $ and $\gamma^{(1)}_{dp} = \gamma^{(2)}_{dp} = 0$). We set the rest of the parameters to be the same as the calculation for temporal filtering, and we choose the integration time window to be $T = \gamma^{-1}$. 

For all three entanglement schemes, we observe a distinct plateau considerably smaller than unity as we narrow down the bandwidth of the spectral filter. In some cases, the entanglement fidelity is even reduced as we decrease the filter bandwidth. This is unexpected, as spectral filtering has been widely employed to enhance photon indistinguishability in solid-state emitters. Our findings suggest that indistinguishability is not the only limiting factor for entanglement fidelity. In fact, in the limit of infinitely narrow filter bandwidth, the spin-tagged photons from each atom after spectral filtering will have perfect indistinguishability, but they will differ by a random overall phase $\phi$. This phase difference will alter the heralded entangled state from the ideal Bell state of $\ket{\Psi^-}=\frac{1}{\sqrt{2}}(\ket{g_1 m_2} - \ket{m_1 g_2})$ to $\ket{\Psi^\phi} = \frac{1}{\sqrt{2}}(\ket{g_1 m_2} - e^{i\phi}\ket{m_1 g_2})$. Although $\ket{\Psi^\phi}$ remains a maximally entangled state, the random fluctuation in the value of $\phi$ among different entanglement attempts will degrade the measured entanglement fidelity.

\section{Comparison of entanglement schemes under optical dephasing and spectral diffusion}
\label{tradeoff}

In this section, we compare the performance of the three different entanglement schemes under the effects of optical dephasing and spectral diffusion. As discussed in Sec. \ref{filtering}, even in the presence of optical dephasing and spectral diffusion, near-unity entanglement fidelity can be achieved through temporal filtering with a sufficiently short time window. Therefore, the entanglement fidelity alone is not adequate for comparing different schemes. Instead, noticing that temporal filtering inherently reduces entanglement efficiency, we will compare the optimal entanglement fidelity at a given entanglement efficiency.

\begin{figure}[t!]
     \centering
         \includegraphics[width=0.45\textwidth]{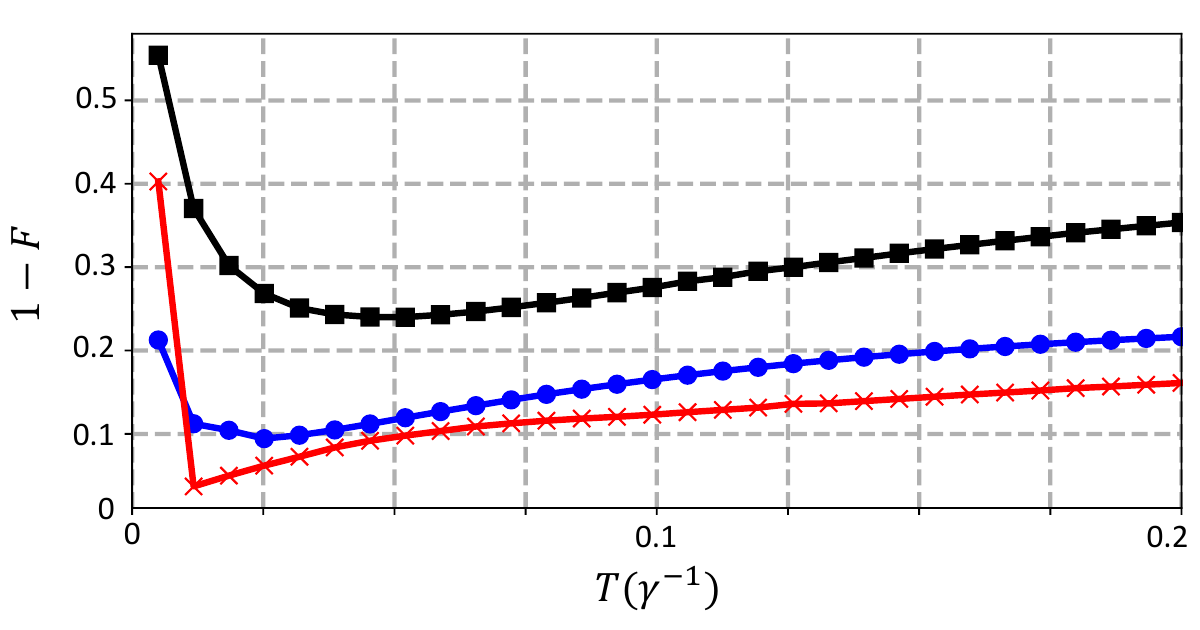}
         \caption{Entanglement infidelity as a function of the integration time window $T$ when the entanglement efficiency is constrained to 1\%, for the spontaneous emission (blue circles), the Raman emission (red crosses), and the resonant scattering scheme (black rectangles).}
         \label{optimizability}
\end{figure}

For each entanglement scheme, we can adjust at least two parameters to achieve the specified entanglement efficiency: the integration time window $T$ and the scheme-dependent parameters that determine the probability of generating a spin-tagged photon. We consider the initial state parameter $\alpha$ for the spontaneous emission scheme, the detuning $\Delta$ and the Rabi frequency $\Omega_{k}$ for the Raman emission scheme, and the amplitude of the coherent field $\beta$ for the resonant scattering scheme. Figure \ref{optimizability} shows the calculated entanglement infidelity as a function of the integration time window $T$ for the spontaneous emission (blue circles), the Raman emission (red crosses), and the resonant scattering scheme (black rectangles), respectively, where the entanglement efficiency is constrained to $1\%$. Here, we define the infidelity as $1-F$. In this calculation, we set $\gamma^{(1)} = \gamma^{(2)} = \gamma $, $\Gamma^{(1)} = \Gamma^{(2)} = 10\gamma $, $\gamma^{(1)}_{dp} = \gamma^{(2)}_{dp} = 5\gamma $, and $\gamma^{(1)}_{sd} = \gamma^{(2)}_{sd} = 0\gamma$. At each integration time window $T$, the scheme-dependent parameters are uniquely determined by optimizing the infidelity under the constraint on the entanglement efficiency.

For all three schemes, we observe that the entanglement infidelity initially decreases as the integration time window $T$ increases, reaches a minimum, and then begins to increase. The increase in entanglement infidelity at large $T$ occurs because temporal filtering becomes less effective in this regime, making each scheme more susceptible to optical dephasing. The initial decrease of infidelity with increasing $T$ can be explained as follows: when $T$ is very short, a high spin-tagged photon generation probability is required to maintain the fixed efficiency of 1\%. For the spontaneous emission and Raman emission schemes, this high photon generation probability increases the likelihood of photon emission from both atoms, thereby increasing entanglement infidelity. For the resonant scattering scheme, a large $\beta$ increases the likelihood of exciting the atom with multi-photon within the excited state lifetime, also leading to increased infidelity.

\begin{figure}[t!]
     \centering
         \includegraphics[width=0.5\textwidth]{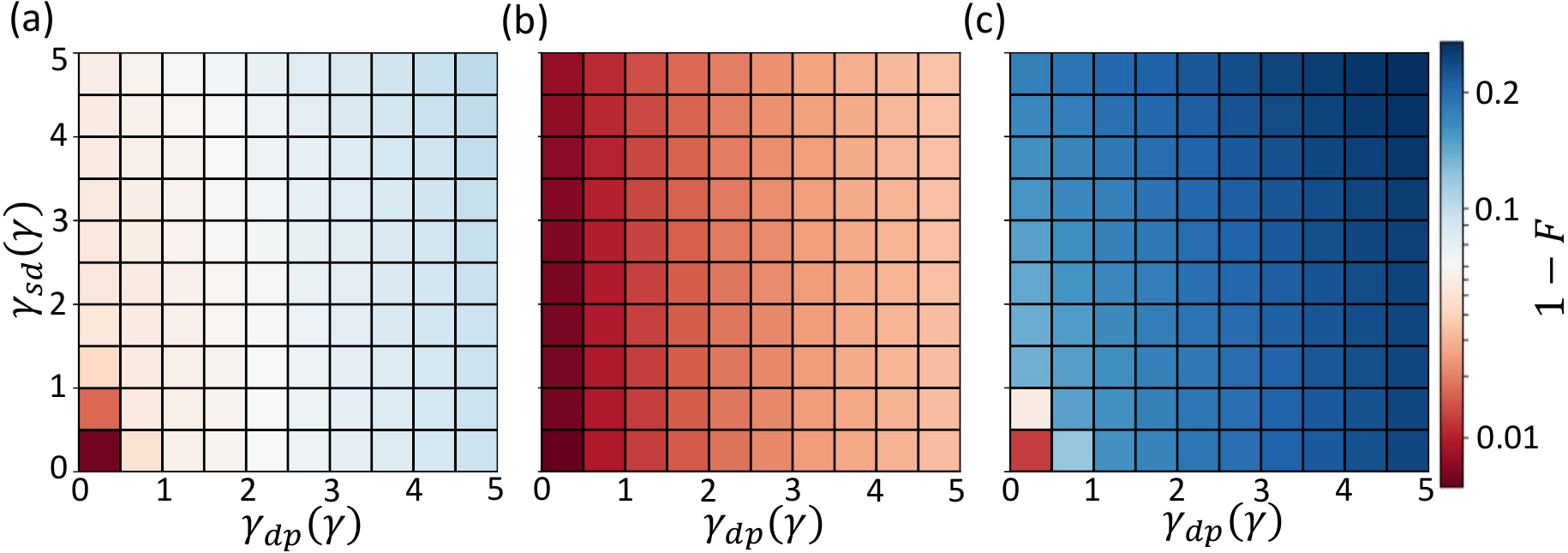}
         \caption{Optimal entanglement infidelity as a function of optical dephasing rate and spectral diffusion linewidth for the spontaneous emission scheme (a), the Raman emission scheme (b), and the resonant scattering scheme (c). In the calculation, we constrain the entanglement efficiency to be 1\%.}
         \label{Fidelity}
\end{figure}

We now compare the optimal performance of the three entanglement schemes under different values of optical dephasing rate and spectral diffusion linewidth. Figure \ref{Fidelity} shows the infidelity as a function of both the optical dephasing rate and the spectral diffusion linewidth for the spontaneous emission scheme (a), the Raman emission scheme (b), and the resonant scattering scheme (c), respectively, when the entanglement efficiency is constrained to 1\%. One may immediately notice that the entanglement infidelity of the Raman emission scheme is remarkably insensitive to the spectral diffusion linewidth. This is because the frequency of the spin-tagged photon generated via Raman emission is determined by the driving laser rather than the atomic transition frequencies. Consequently, even if the optical transition of the atom experiences spectral diffusion, the frequency of the spin-tagged photon remains unaffected. Although fluctuations in the optical transition frequency can alter the probability of Raman emission, these fluctuations are suppressed by a factor of $\gamma_{sd}/\Delta$.

In contrast, the entanglement infidelity of both the spontaneous emission scheme and the resonant scattering scheme increases considerably as the spectral diffusion linewidth increases.  For the spontaneous emission scheme, when there is a detuning $\delta\omega$ between the two atoms, the heralded state upon detecting a photon will become $\ket{\Psi^\phi} = \frac{1}{\sqrt{2}}(\ket{g_1 m_2} - e^{i\phi}\ket{m_1 g_2})$, where $\phi = \delta\omega \cdot t$ is the phase difference of the spin-tagged photons from each atom, and $t$ is the time when the photon is detected. Since $\delta\omega$ varies in each entanglement attempt, the average infidelity increases due to the random fluctuation in the value of $\phi$. The range of fluctuation of $\phi$ is in the order of $\gamma_{sd}T$. For the resonant scattering scheme, when $\delta\omega$ fluctuates, the phase of the scattered photon between the $\ket{g_1m_2}$ and $\ket{m_1g_2}$ components also fluctuates with a range of $\gamma_{sd}T$. In addition, the amplitude of the $\ket{g_1m_2}$ and $\ket{m_1g_2}$ components in the heralded state will also deviate from the ideal value of $1/\sqrt{2}$ because the likelihood of each atom being excited from the ground state $\ket{g}$ is no longer identical. For this reason, the infidelity of the resonant scattering scheme increases faster than the spontaneous emission scheme.

\label{comparison}

The entanglement infidelity of all three schemes increases considerably as we increase the optical dephasing rate. For the spontaneous emission and the Raman emission scheme, the increase in entanglement infidelity results from the reduction in the purity of the spontaneously emitted photons. For the resonant scattering scheme, the increase in entanglement infidelity is due to the random and rapid phase and amplitude fluctuation of the $\ket{g_1m_2}$ and the $\ket{m_1g_2}$ components. The performance of the Raman emission scheme is more resilient against optical dephasing compared with the other two, since the phase fluctuation in a Raman-emitted photon is reduced by a factor $\Omega/\Delta$.

Finally, we note that the resonant scattering scheme has a higher infidelity compared to the other two schemes,  even when both the optical dephasing and spectral diffusion are absent. The infidelity of the resonant scattering scheme is approximately an order of magnitude higher than that of the spontaneous emission scheme, consistent with the findings in Ref. \cite{beukers2024remote}. This result suggests that multi-photon scattering in the resonant scattering scheme degrades fidelity more significantly than two-photon emission in the spontaneous emission scheme. Notably, Ref. \cite{beukers2024remote} reports that replacing the incident weak coherent field with a single photon improves the fidelity of the resonant scattering scheme beyond that of the spontaneous emission scheme, supporting our conjecture.

\begin{figure}[t!]
     \centering
         \includegraphics[width=0.5\textwidth]{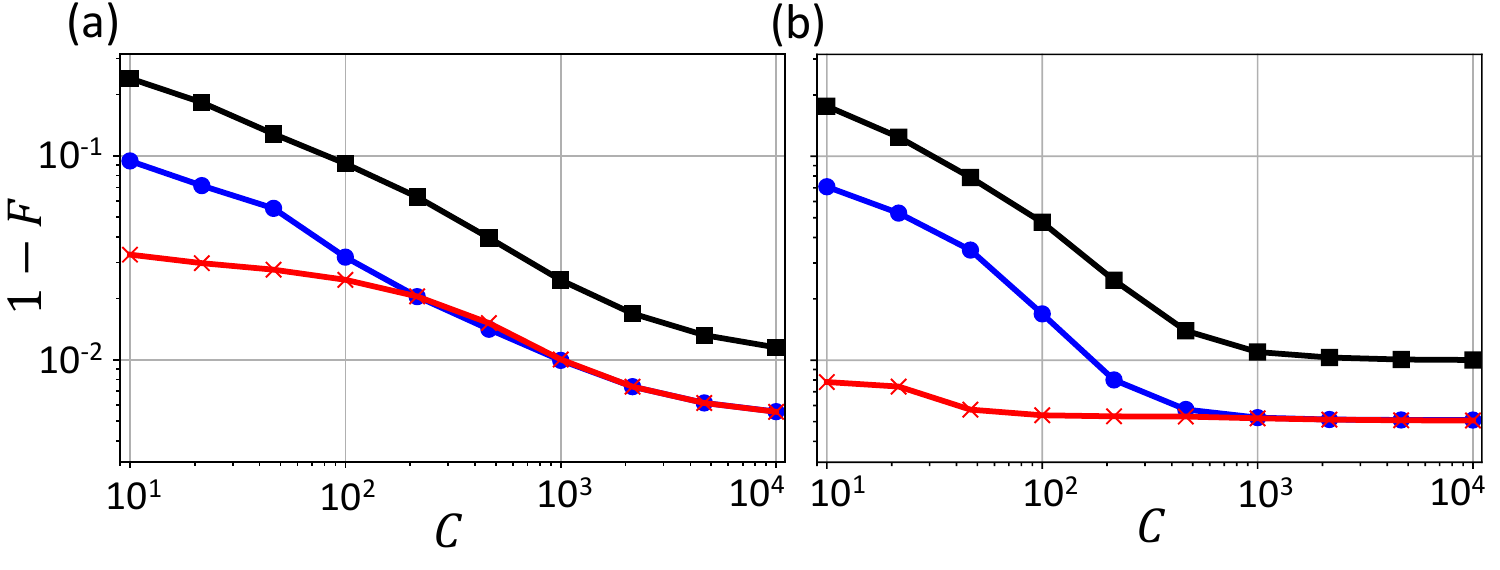}
         \caption{Optimal entanglement infidelity as a function of cooperativity under optical dephasing (a) and spectral diffusion (b) when the entanglement efficiency is constrained to 1$\%$, for the spontaneous emission scheme (blue circles), the Raman emission (red crosses), and the resonant scattering scheme (black rectangles).}
         \label{cooperativity}
\end{figure}

Figure \ref{cooperativity} presents the infidelity as a function of cooperativity, defined as $C = \Gamma/\gamma$, for the spontaneous emission (blue circles), Raman emission (red crosses), and resonant scattering (black rectangles) scheme, respectively, with the entanglement efficiency constrained to 1\%. In Fig. \ref{cooperativity}(a), we set \(\gamma^{(1)}_{dp} = \gamma^{(2)}_{dp} = 5\gamma\) and \(\gamma^{(1)}_{sd} = \gamma^{(2)}_{sd} = 0\). In Fig. \ref{cooperativity}(b), we set \(\gamma^{(1)}_{dp} = \gamma^{(2)}_{dp} = 0\) and \(\gamma^{(1)}_{sd} = \gamma^{(2)}_{sd} = 5\gamma\). The fidelity of all three schemes improves initially as the cooperativity of the spin-photon interface increases, and then plateaus when the fidelity starts to be limited by two-photon emission or multi-photon scattering. When the cooperativity is sufficiently large (on the order of \(10^2\) to \(10^3\)), the spontaneous emission scheme achieves a fidelity comparable to that of the Raman emission scheme. In contrast, the resonant scattering scheme continues to exhibit a higher infidelity. This limitation is again due to undesired multi-photon scattering processes that arise when employing a weak coherent field as the input.

\section{Conclusion}
\label{conclusion}

In this article, we quantitatively compare the performance of three different photon-mediated entanglement generation schemes under the influence of optical dephasing and spectral diffusion, two imperfections particularly relevant for solid-state quantum emitters. We find that the Raman emission scheme generally achieves the best performance in the presence of these imperfections, while the resonant scattering scheme results in infidelity that is approximately an order of magnitude worse than the other two. Our calculations also reveal that temporal filtering is a useful technique for improving achievable entanglement fidelity among the three schemes we consider, whereas spectral filtering has limitations in boosting entanglement fidelity. Our results provide a clear guide on the selection of the optimal photon-mediated entanglement generation schemes and measurement strategies for different solid-state quantum emitters.

We note that our analysis only considers imperfections due to emitter linewidth broadening. In practice, other measurement imperfections may significantly impact the choice of entanglement scheme. For instance, the Raman emission scheme requires excellent phase stability between the lasers driving each atom, which becomes particularly challenging when the atoms are separated by kilometers. The spontaneous emission scheme, on the other hand, demands that the timing precision between the two remote quantum nodes be much better than the excited state lifetime. For emitters with Purcell-enhanced optical transitions, the required timing resolution would need to be at the level of picoseconds, which, while generally easier to achieve than maintaining perfect phase coherence between two remote lasers, still presents significant experimental challenges and complexities in implementation. Conversely, the resonant scattering scheme can be designed such that a single photon sequentially scatters from each atom \cite{knaut2024entanglement}, eliminating the need for distributing timing or phase stability between two remote nodes. Therefore, the ultimate choice of entanglement scheme should be based on the constraints of both the experimental setup and the properties of the quantum emitters. 

\section*{Acknowledgment}
We acknowledge funding from the National Science Foundation (2032567 and 2317149), the Air Force Office of Scientific Research (AFOSR) (FA2386-21-1-4084), and the W. M. Keck Foundation. S.S. acknowledges support from the Sloan Research Fellowship.

\end{document}